\documentstyle[pra,aps]{revtex}
\newcommand{\dfrac}[2]{
                       \displaystyle{\frac{{#1}}{{#2}}}
	                  }

\newcommand{\dsqrt}[1]{\displaystyle{\sqrt{#1}}}
\def\dint{\displaystyle{\int}}

\def\be{\begin{eqnarray}}
\def\ee{\end{eqnarray}}
\def\ba{\begin{array}}
\def\ea{\end{array}}
\def\non{\nonumber\\}

\def\noa{\noalign{\vskip0.4cm}}
\begin{document}
\draft
%%%%%%%%%%%%%%%%%%%%%%%%%%%%%%
\title{Quantum nondemolition measurement of a single electron spin in a quantum dot}
\author{Mitsuro Sugita\cite{byline}, Susumu Machida, Yoshihisa Yamamoto}
\address{
Quantum Entanglement Project, ICORP, JST and 
Edward L. Ginzton Laboratory, Stanford University, Stanford, California 94305
}
%%%%%%%%%%%%%%%%%%%%%%%%%%%%%%
\date{\today}
\maketitle
%%%%%%%%%%%%%%%%%%%%%%%%%%%%%%
\begin{abstract}
We propose a scheme for the quantum nondemolition (QND) measurement of 
a single electron spin in a single quantum dot (QD).
Analytical expressions are obtained for the optical Faraday effect 
between a quantum dot exciton and microcavity field.
The feasibility of the QND measurement of a single electron spin is
discussed for a GaAs/AlAs microcavity with an InAs QD.
%%%%%%%%%%%%%%%%%%%%%%%%%%%%%%
\end{abstract}
%%%%%%%%%%%%%%%%%%%%%%%%%%%%%%
%
% body of paper here
% 
%%%%%%%%%%%%%%%%%%%%%%%%%%%%%%%%%%%%%%%%%%%%%%%%%%%%%%%%%%%%
\section{Introduction}
%%%%%%%%%%%%%%%%%%%%%%%%%%%%%%%%%%%%%%%%%%%%%%%%%%%%%%%%%%%%
%%%%%% background and the needs %%%%
The measurement of a single electron spin in a semiconductor QD
is of great importance,
because an electron spin in a QD is proposed to be a key element
to construct new optical and electronic devices, 
including a spin-based quantum computer \cite{loss1,atac}. 
A long decoherence time of the electronic spin is vital, 
but no experimental result for the lifetime of a truly single electron spin
in a single QD has been reported so far,
due to the lack of capability of measuring a single electron spin
non-destructively.
%The readout of a spin must be also done in the computation itself
%to take out the result \cite{loss1}.

%%%%%% predecessors and the problem %%%%%%
Photoluminescence (PL) measurement has been used to access
an individual QD \cite{heller,toda1}.
Since PL measurement is a destructive process
in a sense that real excitation and recombination of an electron-hole pair is involved,
it is impossible to monitor a single electron spin non-destructively and continuously
for a certain period of time.
An alternative method, 
time-resolved Faraday rotation (TRFR) has been developed \cite{kikkawa1}, 
which enables one to monitor the spin continuously,
but a single spin measurement sensitivity has not been achieved yet \cite{kikkawa2,gupta}.

%%%%%% proposed solution and good points solving the problem %%%%%%%%
In this paper, we propose an optical readout method
based on the interaction between
a single electron spin in a single QD and microcavity field, 
which enables us to perform QND monitoring of a single spin.
In the proposed system, the phase shift of the probe light 
incurred by a single spin is enhanced by a microcavity.
The coupled cavity-QD system is formulated by the quantum Langevin equations,
and the response of the input probe light is analyzed quantitatively. 
It is shown 
that the polarization-sensitive Mach-Zehnder interferometer operating 
at the standard-quantum-limit can perform the QND measurement of a single electron spin
, if the QD is incorporated into a high-Q microcavity or photonic bandgap structure.
 
 This proposal is motivated by the recent experimental demonstration of a GaAs/AlAs DBR
 post-microcavity with a single InAs QD and an optical mode volume of $\sim (\lambda/n)^3$
 \cite{solomon} and theoretical prediction that such a microcavity should have a Q-value of
 higher than $10^4$ \cite{matt}.
%%%%%%%%%%%%%%%%%%%%%%%%%%%%%%%%%%%%%%%%%%%%%%%%%%%%%%%%%%%%
\section{System Configuration}
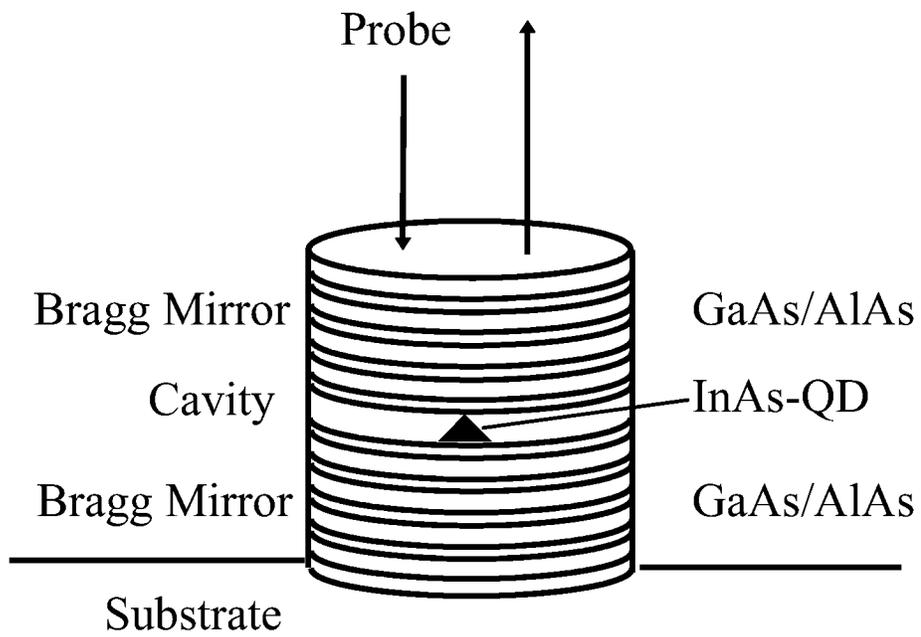
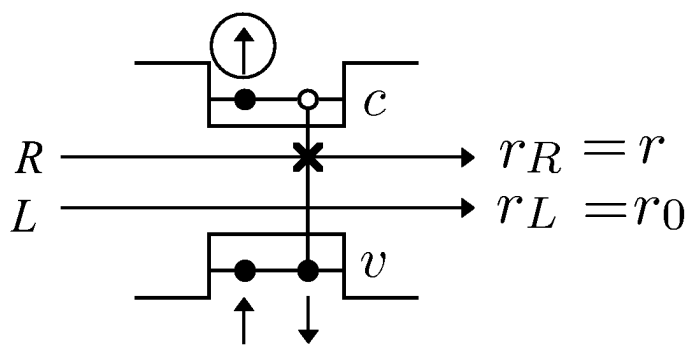
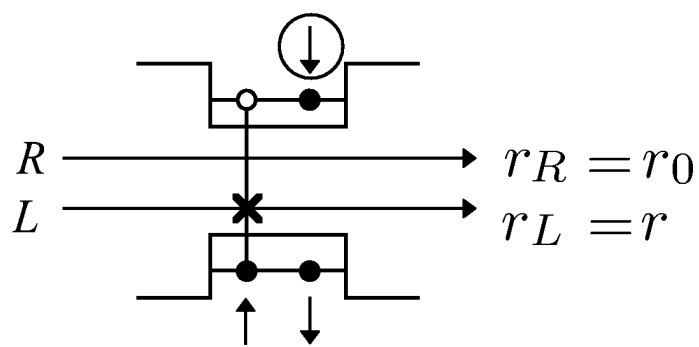
\begin{figure}
\caption{
Sample and its electronic states.
(a) Sample: a single QD in a micropost semiconductor cavity.
Three dimensional confinement is due to the micropost structure
and multilayer Bragg mirrors. The Bragg mirror consists of $\lambda/4$
thick AlGaAs and AlAs layers.
An InAs-QD is embedded in a GaAs cavity. 
%The cavity length is chosen to
%be $m \lambda/(2n)$ (m is an integer, n is the refractive index).
(b),(c) Electronic state configurations of the QD, 
coupling with probe fields:
$c$ and $v$ indicate conduction and valence band ground states of the QD.
Arrows are for the spin states.
Filled states are with black dots, empty states are with white.
$R$ and $L$ denote circular polarizations of the probe fields, which interact
with electron excitation ( indicated by x's ) obeying the selection rule.
$r_R$ and $r_L$ are the reflectivities.
}
\label{qd}
\end{figure}

%%%%%%%%%%%%%%%%%%%%%%%%%%%%%%%%%%%%%%%%%%%%%%%%%%%%%%%%%%%%
Figure \ref{qd} shows essential components of the scheme.
(a) is the sample which consists of an InAs QD and GaAs/AlAs micropost cavity, and
(b) and (c) are the electronic states and their coupling to the probe light (microcavity field).
Suppose incident probe light traveling along z-axis has a linear polarization along x-axis.
The polarization state can be decomposed into right and left circular polarizations, 
which we denote by the subscripts $R$ and $L$ respectively;
%%%%%%%%%%%%%%%%
\be
%%%%%%%%%%%%%%%%
A_{in}	&=&
	\left( \ba{c}
		1 \\
		0
	          \ea
	\right)
	=
	\dfrac{1}{2}
		\left( \ba{c}
			1 \\
			i
	    	      \ea
		\right)
		+
	\dfrac{1}{2}
		\left( \ba{c} 
			1 \\
			-i
	    	      \ea
		\right)
	= A_R + A_L .
\label{2a}
%%%%%%%%%%%%%%%%
\ee
%%%%%%%%%%%%%%%%
If the sample has circular-birefringence,
and the reflection coefficients are $r_R$ and $r_L$, respectively,
the reflected probe light field is
%%%%%%%%%%%%%%%%
\be
%%%%%%%%%%%%%%%%
A_{out}
	&=&
		r_R A_R + r_L A_L
	\non
	&=&
		\dfrac{1}{2}
			\left( \ba{c} 
				r_R + r_L \\
				i (r_R - r_L)
			       \ea
			\right)
	= 
		\left( \ba{c}
			A_x \\
			A_y
		        \ea
		\right) .
\label{2b}
%%%%%%%%%%%%%%%%
\ee
%%%%%%%%%%%%%%%%

In Fig. \ref{qd}(b) and (c),
the four different QD energy states,
%%%%%
ground state electron and hole states with opposite spins,
%%%
and their coupling to probe light with
%%%%%%%
two circular polarizations,
$R$ and $L$, are shown.
The conduction band ground state of the QD is 
the angular momentum state of 
$\pm$ 1/2, and the valence band ground state is that of $\pm$ 3/2,
%%%%%%
if we neglect the heavy hole and light hole band mixing effect.
%%%%%%
Because of the selection rule, which defines allowed and forbidden transitions 
in the electric dipole approximation,
the up-spin channel only couples to $L$ polarization and the down-spin does to
$R$ polarization. 
$r$ and $r_0$ are the responses of the QD system to the probe light
for the allowed and forbidden channels, respectively.

The y-component of the reflected probe light for the two cases
in Fig. \ref{qd}(b) (up-spin) and (c) (down-spin) are
%%%%%%%%%%%%%%%%
\be
%%%%%%%%%%%%%%%%
A_{y}	&=&
	\left\{
		\ba{ll}
			+ \dfrac{i}{2} ( r - r_0 ) 	&\quad \mbox{ for up-spin } \\ \noa	
			- \dfrac{i}{2} ( r - r_0 ) 	&\quad \mbox{ for down-spin }
		\ea
	\right.
\label{2c}
%%%%%%%%%%%%%%%%
\ee
%%%%%%%%%%%%%%%%

If we introduce the amplitude and the phase responses 
for the two polarizations separately,
as $r_{R,L} = | r_{R,L} | e^{ i \theta_{R,L} }$, via
%%%%%%%%%%%%%%%%
\be
%%%%%%%%%%%%%%%%
\bar{r}		&\equiv&	\dfrac{ |r_R| + |r_L| }{2},   
	\quad dr \equiv |r_R| - |r_L|,  
\non\noa
\bar{\theta}	&\equiv&	\dfrac{ \theta_R + \theta_L }{2}, 
	\quad d \theta \equiv \theta_R - \theta_L, 
\label{1d}
%%%%%%%%%%%%%%%%
\ee
%%%%%%%%%%%%%%%%
the reflected probe amplitude is represented in first order of $d r$ and $d \theta$ as,
%%%%%%%%%%%%%%%%
\be
%%%%%%%%%%%%%%%%
A_{out}	&=&	\bar{r} e^{ i \bar{\theta} }
		\left( \ba{c} 
			1 \\
			- \dfrac{ d \theta }{2} - i \dfrac{ d r }{ 2 \bar{r} }
	    	       \ea
		\right) .
\label{2e}
%%%%%%%%%%%%%%%%
\ee
%%%%%%%%%%%%%%%%
We use a dispersive regime with negligible real excitation, so 
%the response is in-phase, i.e.
, $ d \theta \gg d r/ \bar{r} $.
The linearly polarized light is obtained as an output with a slight rotation proportional 
to the difference of the phase between the two circular polarizations. 
This rotation corresponds to a simple magnetooptical Faraday effect by a single spin.
The signal is
%%%%%%%%%%%%%%%%
\be
%%%%%%%%%%%%%%%%
A_{y}	&=&
	\left\{
		\ba{ll}
			+ \dfrac{ \theta - \theta_0 }{2} \bar{r} e^{ i \bar{\theta} }
				&\quad \mbox{ for up-spin } \\ \noa
			- \dfrac{ \theta - \theta_0 }{2} \bar{r} e^{ i \bar{\theta} }
				&\quad \mbox{ for down-spin }
		\ea
	\right. ,
\label{2f}
%%%%%%%%%%%%%%%%
\ee
%%%%%%%%%%%%%%%%
where $ \theta \equiv Arg(r) $ and $ \theta_0 \equiv Arg(r_0) $.

In Fig. \ref{response}, the responses $\theta$ and $\theta_0$, with and without
the QD electron spin, are shown as a function of the probe light frequency
%%%
detuning from the QD transition frequency. 
%%%
One can measure the signals in Eq.(\ref{2f}) using a polarization sensitive
interferometer shown in Fig. \ref{setup}.
Details of the responses and the setup are described in the following sections.

%For simplicity and paying attention to the
%spin measurement, hereafter we mainly discuss about the cases (a) and (b).

%%%%%%%%%%%%%%%%%%%%%%%%%%%%%%%%%%%%%%%%%%%%%%%%%%%%%%%%%%%%
\section{Expected phase rotation}
\label{sec3}
%%%%%%%%%%%%%%%%%%%%%%%%%%%%%%%%%%%%%%%%%%%%%%%%%%%%%%%%%%%%

In this section, we derive the responses $r$ and $r_0$ of an optical microcavity 
with and without photon-exciton coupling.
Figure \ref{coupling} shows a mathematical model, where three systems couple with
each other by the coupling constants, $\gamma_{p}$, $\Omega$, $\gamma_{ex}$.
For simplicity, we assume the cavity is single-sided, 
i.e., the second mirror is perfect.
The reason for introducing a microcavity is that the Faraday effect 
of a single spin would be small and need to be enhanced by the
photon storage effect of the cavity.

The Hamiltonian for our system is
%%%%%%%%%%%%%%%%
\be
%%%%%%%%%%%%%%%%
H	&=&	H_{p} + H_{ex} + H_{res} + H_{p-ex} + H_{p-res},
\label{3a}
%%%%%%%%%%%%%%%%
\ee
%%%%%%%%%%%%%%%%
where
%%%%%%%%%%%%%%%%
\be
%%%%%%%%%%%%%%%%
H_{p} &=&
			\hbar \omega_{p}  \sum_{i} a^{\dagger}_{i}  a_{i}
	,\quad H_{ex} =
			\dfrac{\hbar \omega_{ex}}{2} \sum_{j} \sigma_{z j}
	,\quad H_{res} =
			\dint \hbar \omega \sum_{i} b^{\dagger}_{i}(\omega) b_{i}(\omega) d\omega ,
\non\noa
H_{p-ex} &=&
			i \dfrac{\hbar}{2} \sum_{ij}
			( \Omega_{ij} a_{i} \sigma_{+j} - \Omega^{\ast}_{ij} a^{\dagger}_{i} \sigma_{-j} )
	,\quad H_{p-res} =
			i \hbar \dint \sum_{ii'} \kappa_{ii'} (\omega)
				( a^{\dagger}_{i} b_{i'}(\omega) - a_{i} b^{\dagger}_{i'}(\omega) ) ,
\nonumber
\label{3aa}
%%%%%%%%%%%%%%%%
\ee
%%%%%%%%%%%%%%%%
where
%%%%%%%%%%%%%%%%
\be
%%%%%%%%%%%%%%%%
\Omega_{ij} &=& \dfrac{2e}{\hbar} 
					\dsqrt{ \dfrac{ \hbar \omega_{p}  }
								  { 2 \epsilon V_{cav} }
						  }
					<c j| \dfrac{ x \pm iy}{\dsqrt{2} } |v j>
\quad (\enskip + \enskip \mbox{for} \enskip i=1,
		\enskip - \enskip \mbox{for} \enskip i=2 \enskip) ,
\non\noa
\sigma_{- j}	&=&	d^{\dagger}_{v j} d_{c j},
\enskip
\sigma_{+ j}	=	d^{\dagger}_{c j} d_{v j},
\enskip
\sigma_{z j}	=	d^{\dagger}_{c j} d_{c j} - d^{\dagger}_{v j} d_{v j} ,
%%%%%%%%%%%%%%%%
\ee
%%%%%%%%%%%%%%%%
under the rotating wave and the electric dipole approximations \cite{jaynes,collett,pau}.
Here $i$ indicates the polarization of the photon fields: $i=1$ for $R$ and $i=2$ for $L$.
$j$ indicates the spin of the electron in the QD: $j=1$ for up-spin and $j=2$
for down-spin.
$d$ and $d^{\dagger}$ are the annihilation and creation operator of the electrons
in the QD ($d d^{\dagger} + d^{\dagger} d = 1$), and subscripts $c$ and $v$ indicate
the conduction and the valence band ground states, respectively. 
%$|cj>$ and $|vj>$ are one-electron states.

Hamiltonian (\ref{3a}) does not include the exciton-reservoir coupling.
We assume that excitonic dipole decays with a phenomenological constant 
$\gamma_{ex}$ due to the coupling to
%%%%
photon and phonon reservoirs.
%radiation modes and phonons.
%As for the continuous modes of a reservoir field,
%%%%
Assuming the cavity mirror does not change the polarization
and
following Gardiner and Collett\cite{collett}, we approximate
 $\kappa_{ii'}(\omega) = \{ \gamma_{p}/(2\pi) \}^{1/2} \delta_{ii'}$
and replace the reservoir field amplitude
 $b_{i}(\omega)$, $b^{\dagger}_{i}(\omega)$ with

%%%%%%%%%%%%%%%%
\be
%%%%%%%%%%%%%%%%
a^{in}_{i}(t)	&=&	\dfrac{-1}{ \dsqrt{ 2 \pi} }
					\dint^{}_{} e^{ -i \omega( t-t_0) } b_{0 i}(\omega) d\omega ,
\non\noa
a^{out}_{i}(t)	&=&	\dfrac{1}{ \dsqrt{ 2 \pi} }
					\dint^{}_{} e^{ -i \omega( t-t_1) } b_{1 i}(\omega) d\omega .
\label{3b}
%%%%%%%%%%%%%%%%
\ee
%%%%%%%%%%%%%%%%
Here $b_{0 i}(\omega)$ and $b_{1 i}(\omega)$ are the values of 
$b_{i}(\omega)$ at $t=t_0$ and $t_1 (t_0<t<t_1)$, 
respectively.

After assuming the selection rule for
%%%
the electric dipole transition,
%the electron excitation and the polarization of the cavity photon,
%%%
i.e.,$\Omega_{ij} = \Omega_{i}\delta_{ij}$, 
the Heisenberg equations of motion for the amplitude of the fields are then,
%%%%%%%%%%%%%%%%
\be
%%%%%%%%%%%%%%%%
\left\{
\ba{ll}
	\dfrac{d{a}_{i}}{dt}
		&= -i \omega_{p} a_{i} - \dfrac{ \gamma_{p} }{2} a_{i}
			 -  \dfrac{\Omega^{\ast}_{i}}{2} \sigma_{i}
			 + \dsqrt{ \gamma_{p} } a^{in}_{i}
\\\noa
	\dfrac{d\sigma_{- i}}{dt}
		&= -i \omega_{ex} \sigma_{- i} - \dfrac{ \gamma_{ex} }{2} \sigma_{- i}
			 -  \dfrac{\Omega_{i}}{2} \sigma_{z i} a_{i}
\\\noa
	a^{out}_{i}
		&= - a^{in}_{i} + \dsqrt{ \gamma_{p} } a_{i}
\ea
\right. .
\label{3c}
%%%%%%%%%%%%%%%%
\ee
%%%%%%%%%%%%%%%%
Eqs.(\ref{3c}) consist of two independent sets corresponding to 
($i=1$: $R$, down-spin) and ($i=2$: $L$, up-spin).

The frequency dependent response of the cavity system is obtained from Eq.(\ref{3c})
with linearization by using $<\sigma_{z i}>$.
%%%%%%%%%%%%%%%%
\be
%%%%%%%%%%%%%%%%
r_{i}(\omega)
	&=& \dfrac{ a^{out}_{i}(\omega) }
		      { a^{in}_{i}(\omega)}
	= 
		- \dfrac{
		     ( \omega - \omega_{p}  - i \dfrac{ \gamma_{p}  }{2} )
		     ( \omega - \omega_{ex} + i \dfrac{ \gamma_{ex} }{2} ) 
			 							- \dfrac{|\Omega_{i}|^2 <\sigma_{z i}>}{4}
		    }
		    {
		     ( \omega - \omega_{p}  + i \dfrac{ \gamma_{p}  }{2} )
		     ( \omega - \omega_{ex} + i \dfrac{ \gamma_{ex} }{2} )
			 							- \dfrac{|\Omega_{i}|^2 <\sigma_{z i}>}{4}
		    }.
\label{3e}
%%%%%%%%%%%%%%%%
\ee
%%%%%%%%%%%%%%%%
%In the absence of the coupling between cavity photon and QD exciton,
When $<\sigma_{z i}>=0$, i.e., the conduction and 
valence band ground states of the same spin are both occupied or both empty,
there is no effective coupling between the cavity photon and the QD exciton, and
$r(\omega)$ is reduce to a cold cavity reflectance $r_0(\omega)$ \cite{walls};
%%%%%%%%%%%%%%%%
\be
%%%%%%%%%%%%%%%%
r_0(\omega)
	&=& - \dfrac{
		     ( \omega - \omega_{p}  - i \dfrac{ \gamma_{p} }{2} )
		    }
		    {
		     ( \omega - \omega_{p}  + i \dfrac{ \gamma_{p} }{2} )		     
		    }.
\label{3f}
%%%%%%%%%%%%%%%%
\ee
%%%%%%%%%%%%%%%%
For the case where one valence electron exists with the spin state,
we assume $<\sigma_{z i}> \simeq -1$, 
%i.e., the electron is almost in the valence band, 
that maximizes the optical Faraday effect.
%The solution (\ref{3e}) with this condition is well-understood
%in the following two limits. 
Hereafter, we omit the subscript $i$.

%%%%%%%%%%%%%%%%%%%%%%%%%%%%%%%%%%%%%%%%%%%%%%%%%%%%%%%%%%%%
%\subsection{dispersive limit with large detuning between the exciton and the cavity}
%%%%%%%%%%%%%%%%%%%%%%%%%%%%%%%%%%%%%%%%%%%%%%%%%%%%%%%%%%%%

The solution (\ref{3e}) is reduced to
%%%%%%%%%%%%%%%%
\be
%%%%%%%%%%%%%%%%
r(\omega)
	&=& - \dfrac{
		     ( \omega - \omega^{\prime}_{p}(\omega)  
		      - i \dfrac{ \gamma^{-}_{p}(\omega) }{2} )
		    }
		    {
		     ( \omega - \omega^{\prime}_{p}(\omega)		   
		      + i \dfrac{ \gamma^{+}_{p}(\omega) }{2} )
		    } ,
\non\noa
\omega^{\prime}_{p}(\omega)
	&=&	\omega_{p} + \dfrac{\Omega^2}{4}
						\dfrac{  \omega-\omega_{p} - \Delta}
						      { ( \omega-\omega_{p} - \Delta )^2 
						       + \left( \dfrac{\gamma_{ex} }{2} \right)^2
						      } ,
\non\noa
\gamma^{\pm}_{p}(\omega)	
	&=&	\gamma_{p} \pm \dfrac{\Omega^2}{4}
						\dfrac{  \gamma_{ex} }
						      { ( \omega-\omega_{p} - \Delta )^2 
						       + \left( \dfrac{\gamma_{ex} }{2} \right)^2
						      } .
\label{3h}
%%%%%%%%%%%%%%%%
\ee
%%%%%%%%%%%%%%%%
where $\Delta \equiv \omega_{ex} - \omega_{p}$.
In the limit of large detuning and probe frequency close to the cavity resonance, 
i.e., $ \Delta \gg \gamma_{ex}/2, \Omega \gg \omega - \omega_{p}$,
%and $ \Delta \gg \Omega \gg \omega - \omega_{p}$, 
$\omega^{\prime}_{p}(\omega) $ and $ \gamma^{\pm}_{p}(\omega)$ become constant;
%%%%%%%%%%%%%%%%
\be
%%%%%%%%%%%%%%%%
	\omega^{\prime}_{p}	&\simeq& \omega_{p} - \dfrac{\Omega^2}{4 \Delta}
,\quad 
	\gamma^{\pm}_{p}	\simeq \gamma_{p} ,
\label{3i}
%%%%%%%%%%%%%%%%
\ee
%%%%%%%%%%%%%%%%
so that the response is the same as that of the empty cavity 
except for a finite frequency shift
$d \omega_{p} = \Omega^2 / ( 4 \Delta )$.

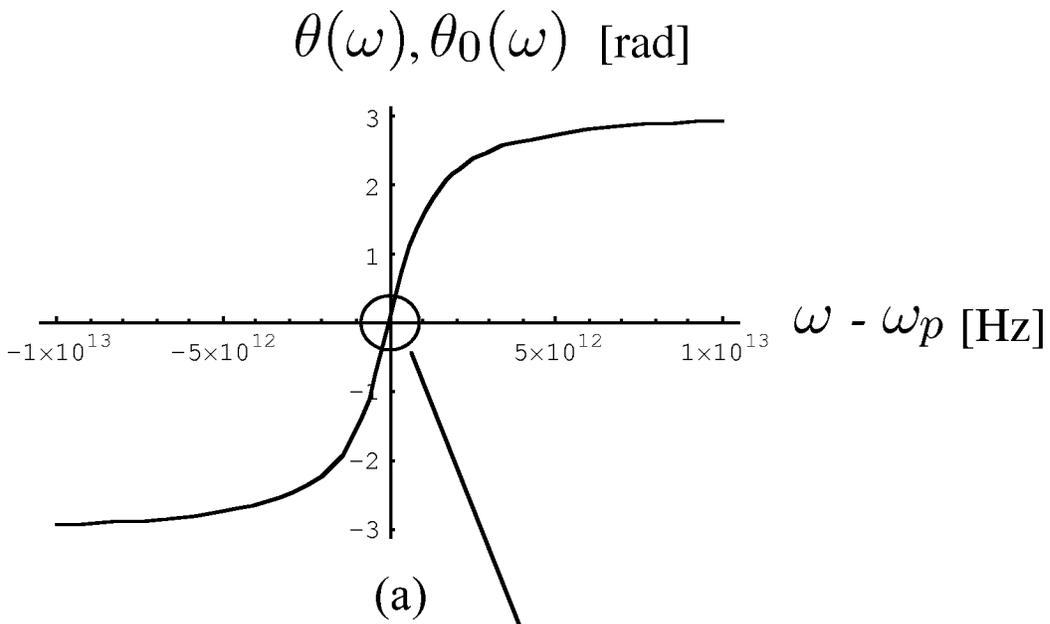
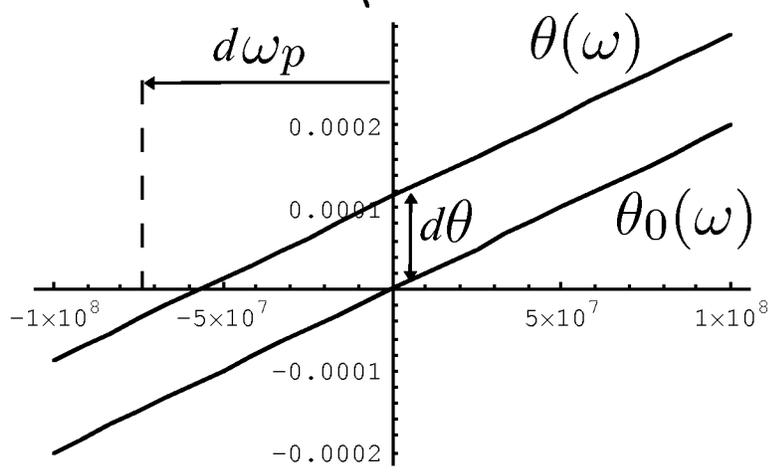
\begin{figure}
\caption{
Phase response of the cavity with and without coupling
to the QD exciton.
(a) Two curves are not distinguished in this scale.
A cold cavity is in phase on the resonance $\omega=\omega_{p}$,
and out of phase off the resonance.
(b) Near the cavity resonance. QD exciton effect appears
as a shift of the resonant frequency by $d \omega_{p}$.
}
\label{response}
\end{figure}

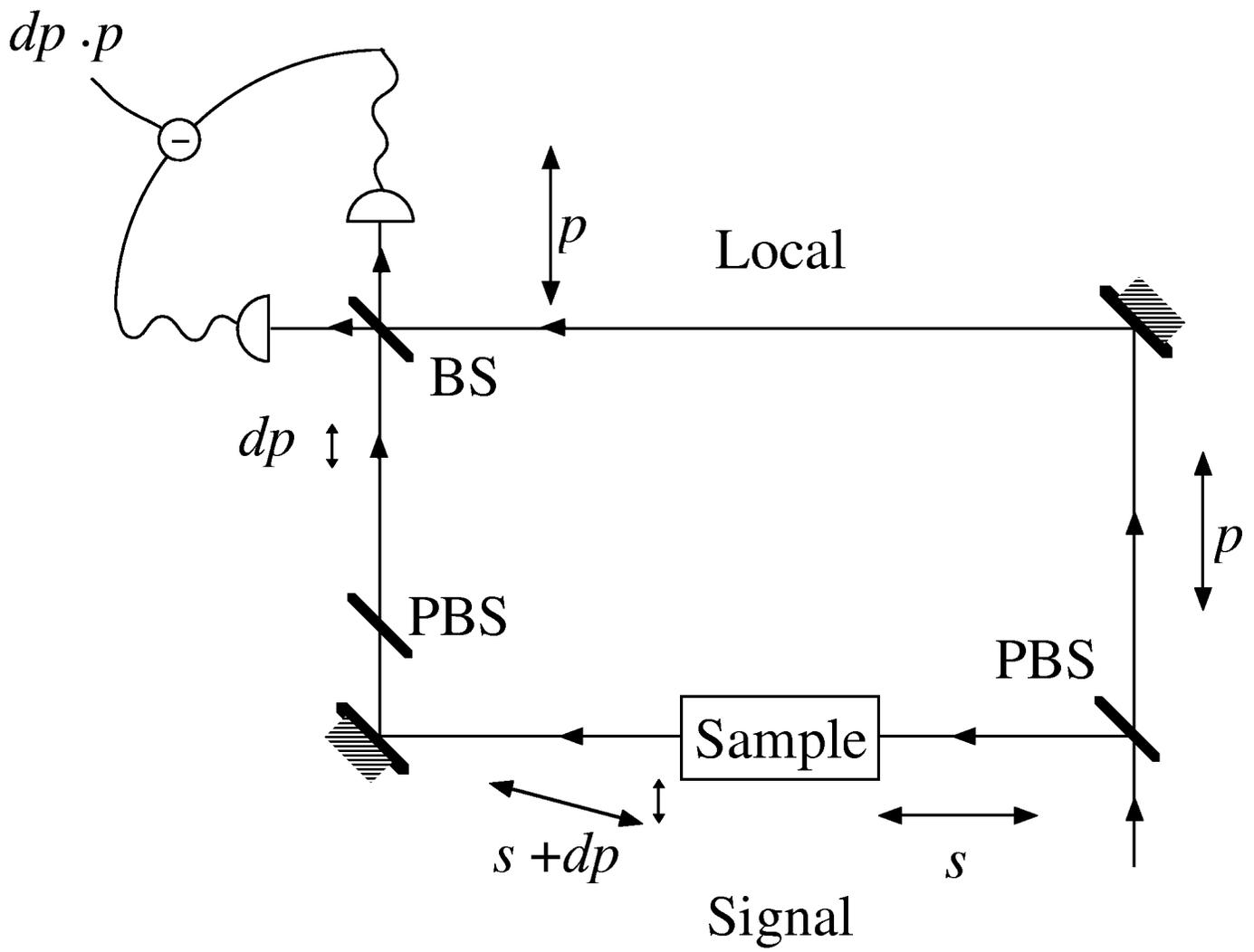
\begin{figure}
\caption{
Proposed experimental setup.
Essential elements of the interferometer.
Laser beam is prepared as 45 degree linear polarization (not shown).
It is divided into p-polarization (for local) and s-polarization (for signal)
by the first polarizing beam splitter (PBS).
After reflected by the sample, signal polarization is rotated by the optical Faraday effect
to have p-polarization ( $d p$ in the figure). From the reflected signal
light, s-polarization is removed by the second PBS and the residual $d p$
is mixed by the beam splitter (BS). The interference between local and signal light
is detected by a balanced homodyne detector.
}
\label{setup}
\end{figure}

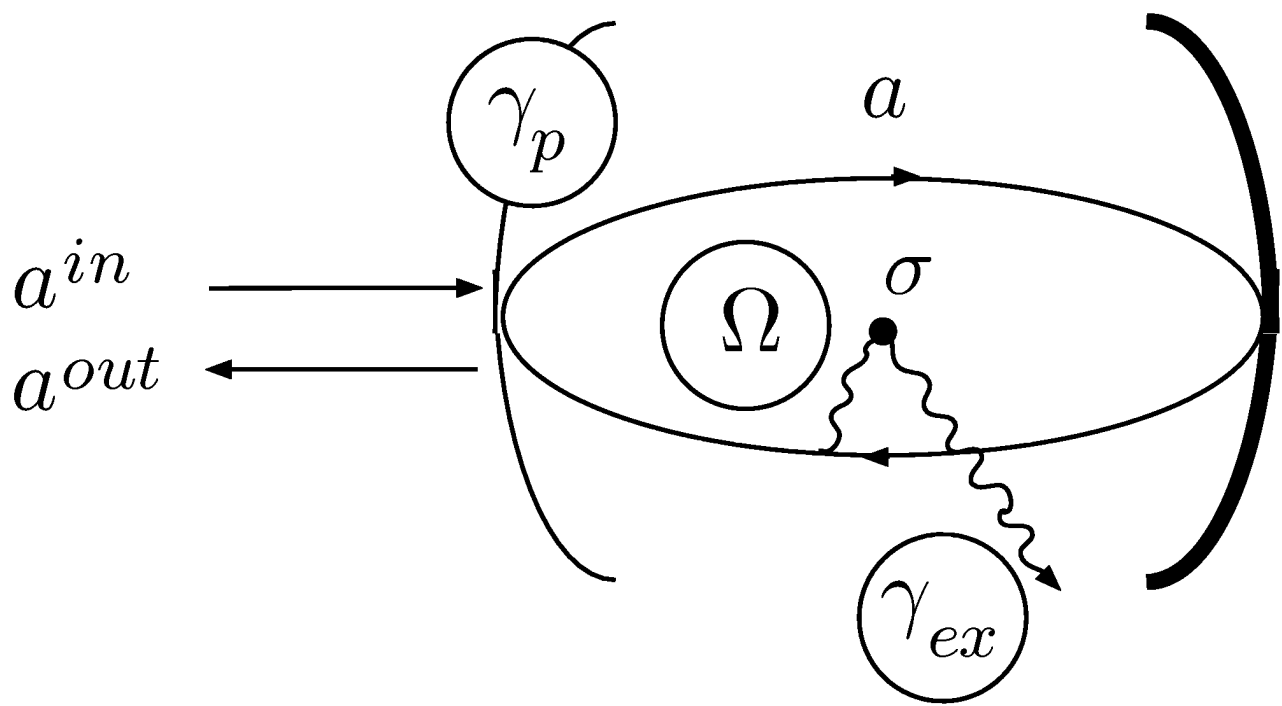
\begin{figure}
\caption{
Fields and couplings of the system.
$a$ for the cavity photon, $\sigma$ for the QD exciton,
and $a^{in}$ and $a^{out}$ for the probe photon. 
The probe photons are coupled with cavity photons by $\gamma_p$
the coupling and dissipation constant. The QD excitons are
coupled with the cavity photons by the Rabi Frequency $\Omega$,
and with the non-cavity modes with dissipation constant  $\gamma_{ex}.$
}
\label{coupling}
\end{figure}

Figure \ref{response} shows the phase responses of the cavity
for the empty and the coupled exciton-field cases. 
The detuning $\Delta =400$THz for the latter case.
Other parameters are 
 $\gamma_{ex} = 10$GHz,
 $\omega_{ex} = 2.1 \times 10^{15}$Hz, 
and 
 $\gamma_{p} = 2$THz ($Q=10^{3}$).
%%%%%%%%%%%%%%%%%%%%
% Rabi freq.
%%%%%%%%%%%%%%%%%%%%
The vacuum Rabi frequency $\Omega$ is evaluated from the spontaneous
emission decay rate formula 
$\gamma_{rad} = \omega^3 e^2|r_{cv}|^2 n_0 /( 3 \pi \epsilon_0 \hbar c^3 )$ via 
%%%%%%%%%%%%%%%%
\be
%%%%%%%%%%%%%%%%
\Omega^2	
		&=& 4 \dfrac{ e^2|r_{cv}|^2 }{ 3 \hbar^2 }
%		      \left( 
		              \dfrac{ \hbar \omega }
		                    { 2 \epsilon_0 n_0^2 V_{cav} }
%		      \right)
		= \dfrac{ 2 \pi c^3}
		          { n_0^3 V_{cav} \omega_{p}^2}
		          \gamma_{rad}.
\label{3j}
%%%%%%%%%%%%%%%%
\ee
%%%%%%%%%%%%%%%%
%where $|c>$ and $|v>$ are assumed to have spherical harmonics $Y^{0}_{0}$
%and $Y^{\pm 1}_{1}$.
Using an exciton radiative recombination rate $\gamma_{rad}=2$GHz, 
cavity volume $V_{cav} \sim (\lambda/n)^3 \sim 2\times 10^{-2} (\mu$m$)^3$,
and $n_0 = 3.6$, $\Omega$ is estimated to be about 300GHz. 
%%%%%%%%%%%%%%%%%%%%%%%%%%%%%%%%%%%%%%%%%%%%%%%%%%%%%

In Fig. \ref{response}, near the cavity resonance,
the effect of exciton appears as the lateral shift of the response curve as mentioned above.
The phase shift is given by
%%%%%%%%%%%%%%
\be
%%%%%%%%%%%%%%
d \theta	&=&
d \omega_{p} \dfrac{4Q}{\omega_p} \quad = \quad \dfrac{\Omega^2}{\Delta \omega_{p}} Q .
\label{3k}
%%%%%%%%%%%%%%
\ee
%%%%%%%%%%%%%%
This is the enhancement that we expect: if the cavity $Q$ is large, the curve is steep and the
phase difference becomes large. 

%%%%%%%%%%%%%

%%%%%%%%%%%%%%%%%%%%%%%%%%%%%%%%%%%%%%%%%%%%%%%%%%%%%%%%%%%%
\section{Feasibility of a QND measurement}
%%%%%%%%%%%%%%%%%%%%%%%%%%%%%%%%%%%%%%%%%%%%%%%%%%%%%%%%%%%%

In this section, 
we discuss the feasibility of the QND measurement
of a single electron spin.
%%%%%%%%
%three conditions
%%%%%%%%
We impose four conditions for the QND measurement for this system:
(A) an expected signal is larger than the shot noise, (B) real excitation of the QD exciton
is negligible during the measurement, (C) integration time of the measurement
is longer than the cavity lifetime, and then (D) spin lifetime of a QD is longer than
the integration time. We will derive conditions (A)-(C) and discuss the region where
the solutions exist, and then compare it with condition (D).

%%%%%%%%%%%%%%%%%%%%%
%condition (A) shot noise
%%%%%%%%%%%%%%%%%%%%%
For the polarization-sensitive interferometer 
with a balanced homodyne detector in Fig. \ref{setup}, the signal is
%%%%%%%%%%%%%%%%
\be
%%%%%%%%%%%%%%%%
|A_y|	&=& \sin({\dfrac{ d \theta }{2}}) \bar{r} |A_{in}| ,
\label{4a}
%%%%%%%%%%%%%%%%
\ee
%%%%%%%%%%%%%%%%
where $|A_{in}|^2 = N_{in}$ is normalized to the incident photon flux of the probe. 
If the probe power and the local oscillator power are equal, 
the signal (normalized to photon flux) due to the single electron spin $N_{SS}$ is
%%%%%%%%%%%%%%%%
\be
%%%%%%%%%%%%%%%%
N_{SS}	&=& N_{in} \sin({\dfrac{ d \theta }{2}}) \bar{r}.
\label{4b}
%%%%%%%%%%%%%%%%
\ee
%%%%%%%%%%%%%%%%
The standard quantum limit (SQL) $N_{SQL}$ for the homodyne detection is given by
%%%%%%%%%%%%%%%%
\be
%%%%%%%%%%%%%%%%
N_{SQL}	&=& \dsqrt{ \dfrac{ N_{in} }{ \tau } },
\label{4bb}
%%%%%%%%%%%%%%%%
\ee
%%%%%%%%%%%%%%%%
where $\tau$ is the integration time.
The SQL originates from the vacuum fluctuation
%%%%
%of the field coming through the signal port of the mixing beamsplitter. 
in the p-polarization incident from the open port of the polarization beam
splitter (PBS) of Fig.3.
%%%%
In order to observe the single spin effect, the signal $N_{SS}$ must be 
greater than $N_{SQL}$, which leads to condition (A):
%%%%%%%%%%%%%%%%
\be
%%%%%%%%%%%%%%%%
N_{in} \tau &>&
\left( |r| \sin\left(\dfrac{d\theta}{2}\right)\right)^{-2},
\label{cond-a}
%%%%%%%%%%%%%%%%
\ee
%%%%%%%%%%%%%%%%
which places a lower bound for the number of photons used in the measurement. 

%%%%%%%%%%%%%%%%%%%%%
%condition (B) real excitation
%%%%%%%%%%%%%%%%%%%%%
Condition (B) is expressed by
%%%%%%%%%%%%%%%%
\be
%%%%%%%%%%%%%%%%
(1-|r|^2) N_{in} \tau &<& 1,
\label{cond-b}
%%%%%%%%%%%%%%%%
\ee
%%%%%%%%%%%%%%%%
which places an upper bound for the number of photons
used during the measurement.

%%%%%%%%%%%%%%%%%%%%%
%Fig.5, N vs \Delta, \gamma_p
%%%%%%%%%%%%%%%%%%%%%
With condition (C) $\tau > 1/\gamma_p$,  the region
where the solutions exit is shown in Fig. 5.
Three curves are plotted, in Fig.5(a), as a function of normalized
detuning, 
using $P=\hbar \omega N_{in}=10$mW, $\omega=2.1\times10^{15}$Hz, and 
$\gamma_p= 0.1 \Omega^2/(4\gamma_{ex})=230$GHz$(Q\sim10^4)$.
The hatched region (top right part) is where the solutions exist.
Figure.5(b) shows the similar result but  $\gamma_p$ is set equal to
$\Omega^2/(4\gamma_{ex})=2.3$THz$(Q\sim10^3)$.
The curves (A) and (B) coincide with each other and 
this is the limit for the solutions to exist: 
if $\gamma_p > \Omega^2/(4\gamma_{ex})$, 
there is no solution. 
We see that the detuning $\Delta/(\gamma_{ex}/2)$ must be
much larger than 1, in this case $\sim10^3$, for the system to have a solution.

%%%%%%%%%%%%%%%%%%%%%%%%%%%%%%%%%%%%%
%far detuning 
%%%%%%%%%%%%%%%%%%%%%%%%%%%%%%%%%%%%%
Figure 6 shows the integration time necessary for the measurement
under the large detuning, as a function of cavity Q value. 
%%%%%%%%%%%%%%%%%%%%%%%%%%%%%%%%%%%%%%%%%%%%%%
The expectation value $<n_{ex}>=<\sigma_{+} \sigma_{-}>$
is used as a parameter, since it  expresses the fundamental limit
$<n_{ex}> \ll 1$ corresponding to the approximation 
$<\sigma_z> \simeq -1$
appeared in Sec. \ref{sec3}.
With the large detuning,
$<n_{ex}>$is derived from Eq.(\ref{3c});
%%%%%%%%%%%
\be
%%%%%%%%%%%
<n_{ex}> 
		&=& \dfrac{\Omega^2}{4\Delta^2} <a^{\dagger} a>
\non\noa
		&=& \dfrac{\Omega^2 Q}{\Delta^2 \omega_p} <a^{in \dagger} a^{in}>
		= \dfrac{\Omega^2 Q}{\Delta^2 \omega_p} N_{in}.
\label{3p}
%%%%%%%%%%
\ee
%%%%%%%%%%
Using Eqs.(\ref{3p}) and (\ref{3k}),  conditions (A)-(C) are
%%%%%%%%%%%
\be
%%%%%%%%%%%
\tau 
		&>& \dfrac{4 \omega_p}{\Omega^2} \dfrac{1}{<n_{ex}>} \dfrac{1}{Q},
\label{3q-a}
\\
\tau 
		&<& \dfrac{1}{\gamma_{ex} <n_{ex}>},
\label{3q-b}
\\
\tau 
		&>& \dfrac{1}{\omega_{p}} Q,
\label{3q-c}
%%%%%%%%%%
\ee
%%%%%%%%%%
respectively. 
Eq.(\ref{3q-b}) reveals that the real excitation rate is the product of
the expectation value of the exciton number and the exciton's dephasing rate. 
In Fig.6, the vertical solid line is the limit $\gamma_p = \Omega^2/(4\gamma_{ex})$
discussed above. 
%%%
The QND conditions are satisfied only in the right of this vertical line.
%%%
Another solid line expresses Eq.(\ref{3q-c}).
 Three combinations of Eqs.(\ref{3q-a}), (\ref{3q-b}), and (\ref{3q-c}) are shown
 in the figure with
 the parameter $<n_{ex}>$ set to 0.1, 0.01, and 0.001, and corresponding three
 triangular regions are seen with heavy hatching.
 Lightly hatched  part is the total region where the solutions exist. 
We see that in order to achieve the QND measurement,
we need cavity Q of $10^3 \sim 10^4$ and
integration time of about 100ps$\sim$10ns.

Since spin lifetime of a doped-QD exciton is expected to be longer than 100ns \cite{atac,kikkawa2},
the integration time we derived is encouraging to perform the QND measurement. 
It is also suggested that a reasonably high-Q cavities, which are within the reach of experimental
microcavity and photonic bandgap structure, can meet the requirements.

%%%%%%%%%%%%%%%%%%%%%%%%%%%%%%%%%%%%%%%%%%%%%%

%%%%%%%%%%%%%%%%%%%%%%%%%%%%%%%%%%%%%%%%%%%%%%%%%%%%%%%%%%%%
\section{Conclusion}
%%%%%%%%%%%%%%%%%%%%%%%%%%%%%%%%%%%%%%%%%%%%%%%%%%%%%%%%%%%%

We have formulated the response of optical cavity with a single QD using Langevin equations
including both QD exciton and external probe field.
The signal to noise ratio in detecting the spin is evaluated.
We conclude that the QND measurement for a single electron spin in a doped-QD
can be achieved by the proposed scheme with a reasonably high-Q cavity under large
detuning condition.

%%%%%%%%%%%%%%%%%%%%%%%%%%%%%%%%%%%%%%%%%%%%%%%%%%%%%%%%%%%%
%acknowledge
%%%%%%%%%%%%%%%%%%%%%%%%%%%%%%%%%%%%%%%%%%%%%%%%%%%%%%%%%%%%
%\acknowledgements
%
%Encouragement and financial support to M.S. from Canon Inc. are gratefully acknowledged.

%%%%%%%%%%%%%%%%%%%%%%%%%%%%%%%%%%%%%%%%%%%%%%%%%%%%%%%%%%%%
% now the references. delete or change fake bibitem. delete next three
%   lines and directly read in your .bbl file if you use bibtex.
%%%%%%%%%%%%%%%%%%%%%%%%%%%%%%%%%%%%%%%%%%%%%%%%%%%%%%%%%%%%
%References
%%%%%%%%%%%%%%%%%%%%%%%%%%%%%%%%%%%%%%%%%%%%%%%%%%%%%%%%%%%%
%\begin{thebibliography}{99}

%\end{thebibliography}
% \begin{figure}
% \caption{}
% \label{}
% \end{figure}
%%%%%%%%%%%%%%%%%%%%%%%%%%%%%%%%%%%%%%%%%%%%%%%%%%%%%%%%%%%%
%Figures
%%%%%%%%%%%%%%%%%%%%%%%%%%%%%%%%%%%%%%%%%%%%%%%%%%%%%%%%%%%%
%%%%%%%%%
%%%%%%%%%
%%%%%%%%%
%%%%%%%%%%%%%%%%%%
%%%%%%%%%
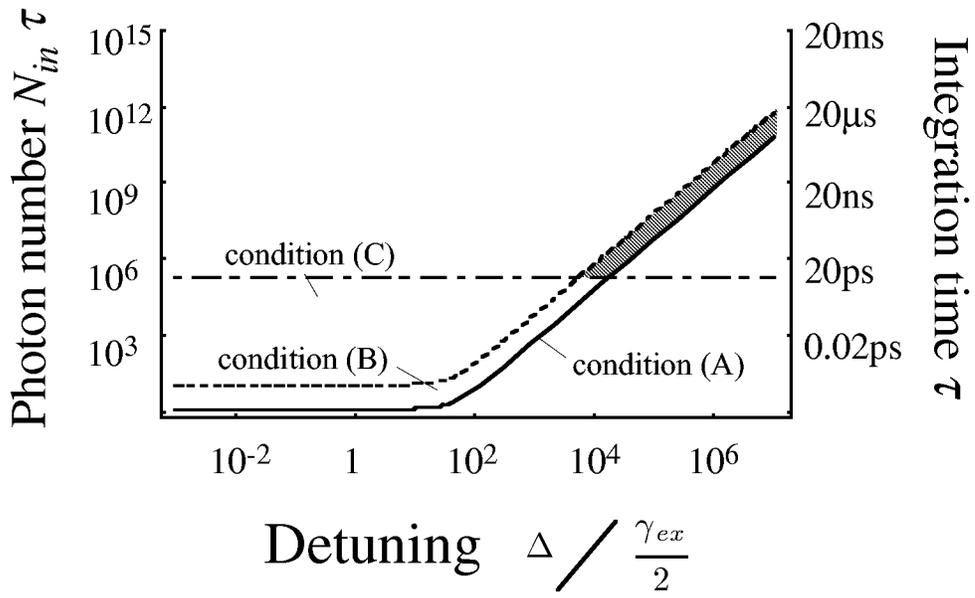
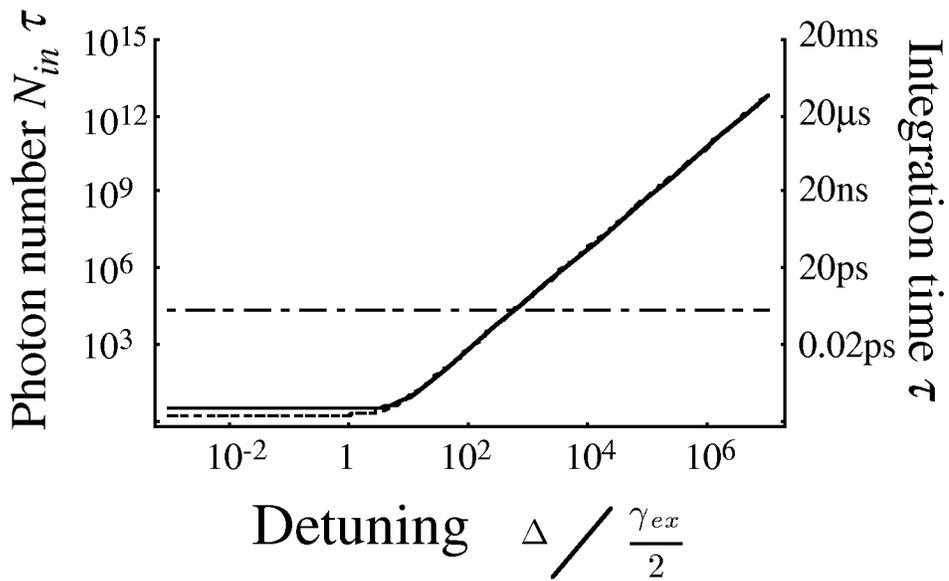
\begin{figure}
\caption{
Conditions for the measurement photon number and the detuning.
Condition (A): the shot noise limitation (solid line) and
condition (B): the real excitation limitation (dotted line) 
have their corner points 
at the detuning close to the exciton dissipation $\gamma_{ex}/2$.
Condition (C): the integration time limitation is constant. 
In (a), the area satisfying all the conditions is hatched (top right).
In (b), conditions (A) and (B) shares the line and the satisfied area is critical. 
}
\label{num}
\end{figure}
%%%%%%%%
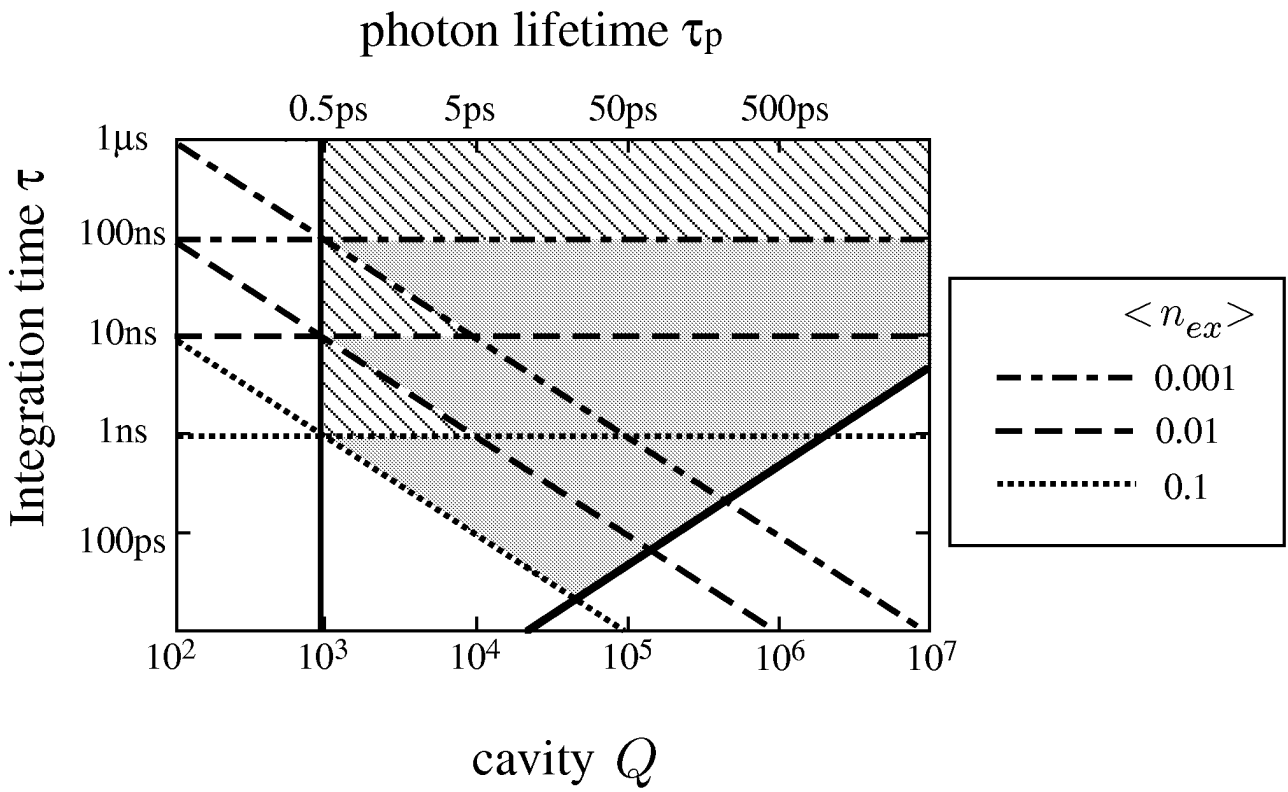
\begin{figure}
\caption{
Conditions for measurement time and cavity Q.
In the legend, the lines used for conditions (A) and (B) with parameter
$<n_{ex}>$ are displayed. Condition (B) appears as horizontal lines.
Condition (C) (solid line from bottom left to top right) and a combination
of conditions (A) and (B) enclose a triangle (dark hatched), which is the
region the solutions exist.
The vertical line shows $\gamma_p > \Omega^2/(2 \gamma_{ex})$, which
the top left apex of the triangular region traces as $<n_{ex}>$ changes.
}
\label{time}
\end{figure}
% \begin{table}
% \caption{}
% \label{}
% \end{tabular}
% \end{table}
%%%%%%%%%%%%%%%%%%%%%%%%%%%%%%%%%%%%%%%%%%%%%%%%%%%%%%%%%%%%
%Tables
%%%%%%%%%%%%%%%%%%%%%%%%%%%%%%%%%%%%%%%%%%%%%%%%%%%%%%%%%%%%

%%%%%%%%%%%%%%%%%%%%%%%%%%%%%%%%%%%%%%%%%%%%%%%%%%%%%%%%%%%%%%

\begin{references}
%%%%%%%%%%%%%%%%%
%\bibitem[*]{byline}E-mail: sugita.mitsuro@canon.co.jp, present address: 
%Optical Technology Research Center, Canon Inc., 23-10 Kiyohara-Kogyodanchi,
%Utsunomiya, Tochigi, Japan 321-3231. 
\bibitem[*]{byline}
Visiting Researcher for 1999-2001 and also with Canon Inc.
Present address: Optical Technology Research Center, Canon Inc., 23-10 Kiyohara-Kogyodanchi,
Utsunomiya, Tochigi, Japan 321-3231. E-mail: sugita.mitsuro@canon.co.jp, 
%%%%%% QC by QD spin %%%%
\bibitem{loss1} D. Loss and D. P. DiVincenzo, 
Phys. Rev. A {\bf 57}, 120 (1998).
%%%%%%%%%%%%%%%%%
%%%%%%% 1 micro sec doped-QD (expect) %%%
\bibitem{atac} A. Imamo\=glu, D. D. Awschalom, G. Burkard, D. P. DiVincenzo,
D. Loss, M. Sherwin, and A. Small,
%, G. Burkard, D. P. DiVincenzo, D. Loss, 
%				M. Sherwin, and A. Small,
%quant-ph/9904096 (1999).
Phys. Rev. Lett. {\bf 83}, 4204 (1999).
%%%%%%%%%%%%%%%%%
%%%%%% single PL %%%%
%\bibitem{martzin} W. Heller and U. Bockelmann, 
%Phys. Rev. B {\bf 55}, 4871 (1996).
%%%%%%%%%%%%%%%%%
%%%%%% single PL %%%%
\bibitem{heller} W. Heller and U. Bockelmann, 
Phys. Rev. B {\bf 55}, 4871 (1996).
%%%%%%%%%%%%%%%%%
%%%%%% Single PL, 1/2,2/3 QD InAs %%%%
\bibitem{toda1} Y. Toda, S. Shinomori, K. Suzuki and Y. Arakawa, 
Phys. Rev. B {\bf 58}, 10147 (1998).
%%%%%%%%%%%%%%%%%
%
%%%%%%%%%%%%%%%%%
%%%%%%% 4ns +2DEG(doped)-QW ZnSe %%%
\bibitem{kikkawa1} J. M. Kikkawa, I. P. Smorchkova, N. Samarth and D. D. Awschalom, 
Science {\bf 277}, 1284 (1997).
%%%%%%%%%%%%%%%%%
%%%%%%% 10ns, 130ns doped-QW GaAs %%%
\bibitem{kikkawa2} J. M. Kikkawa and D. D. Awschalom, Phys. Rev. Lett. {\bf 80}, 4313 (1998).
%%%%%%%%%%%%%%%%%
%%%%%%% 3ns undoped-QD CdSe %%%
\bibitem{gupta} J. A. Gupta, D. D. Awschalom, X. Peng and A. P. Alivisatos, 
				Phys. Rev. B. {\bf 59}, 10421 (1999).
%%%%%%%%%%%%%%%%%
%
%%%%%%% micropost & QD %%%
\bibitem{solomon} G. S. Solomon, M. Pelton and Y. Yamamoto, 
				Phys. Rev. Lett. {\bf 86}, 3903 (2001).
%%%%%%%%%%%%%%%%%
%%%%%%% theoretical limit of microcavity %%%
\bibitem{matt} M. Pelton, Y. Yamamoto,  J. Vuckovic and A. Scherer,
				submitted for publication.
%%%%%%%%%%%%%%%%%
%
%%%%%%%%%%%%%%%%%
%%%%%% exciton coupling %%%%
\bibitem{jaynes} F. W. Cummings, 
Phys. Rev. {\bf 140}, A105 (1965).
%%%%%%%%%%%%%%%%%
%%%%%% in/out coupling %%%%
\bibitem{collett} C. W. Gardiner and M. J. Collett,
Phys. Rev. A {\bf 31}, 3761 (1985).
%%%%%%%%%%%%%%%%%
%%%%%% exciton coupling %%%%
\bibitem{pau} S. Pau et al.,
%, G. Bj\"ork, J. Jacobson, H. Cao, and Y. Yamamoto,
Phys. Rev. B {\bf 51}, 14437 (1995).
%%%%%%%%%%%%%%%%%
%%%%%% in/out response %%%%
\bibitem{walls} D. F. Walls and G. J. Milburn,
{\it Quantum Optics}, 121 (Springer-Verlag, Berlin, 1994).
%%%%%%%%%%%%%%%%%
%%%%%%% 50ps undoped-QW GaAs %%%
\bibitem{barad} S. Bar-Ad and I. Bar-Joseph, Phys. Rev. Lett.{\bf 68}, 349 (1992).
%%%%%%%%%%%%%%%%%
%%%%%%% 50ps undoped-QW GaAs %%%
\bibitem{damen} T. C. Damen, Karl Leo, Jagdeep Shah, and J. E. Cunningham, Appl. Phys. Lett.
				{\bf 58}, 1902 (1991).
%%%%%%% 900ps undoped-QD GaAs %%%
\bibitem{gotoh} H. Gotoh et al.,
%, H. Ando, H. Kamada, A. Chavez-Pirson, and J. Temmyo, 
Appl. Phys. Lett. {\bf 72}, 1341 (1998).
%%%%%%%%%%%%%%%%%
%%%%%%% 200ps undoped-QD InGaAs %%%
\bibitem{kalevich} V. K. Kalevich et al.,
%, M. N. Tkachuk, P. Le Jeune, X. Marie, and T. Amand, 
Physics of the Solid State {\bf 41}, 789 (1999).
%%%%%%%%%%%%%%%%%
\end{references}
\end{document}